\documentclass[runningheads]{svmult}

\usepackage{makeidx}   
\usepackage{graphicx}  
\usepackage{subeqnar}  
\usepackage{multicol}  
\usepackage{physprbb}  
\makeindex             

\begin{document}

\title*{Synthetic Line Profiles for Pulsating Red Giants\footnote{This work was
supported by FWF-project P14365--PHY.}}
\toctitle{Synthetic Line Profiles for Pulsating Red Giants}
\titlerunning{Synthetic Line Profiles for Pulsating Red Giants}

\author{Walter Nowotny\inst{1}
\and Bernhard Aringer\inst{2}
\and Susanne H\"ofner\inst{3}
\and Rita Gautschy-Loidl\inst{4}
\and Josef Hron\inst{1}
\and Walter Windsteig\inst{1}}

\authorrunning{Walter Nowotny et al.}

\institute{Institut f\"ur Astronomie der Universit\"at Wien, 
           Vienna, Austria
\and Instituut voor Sterrenkunde, Leuven, Belgium
\and Dept. of Astronomy and Space Physics, Uppsala University, Uppsala, Sweden
\and Basel, Switzerland}

\maketitle              

\begin{abstract}
Pulsation influences atmospheric structures of variable AGB stars (Miras)
considerably. Spectral lines of the CO $\Delta\,v$=3 vibration--rotation bands
(at $\approx$1.6\,$\mu$m) therefore have a very characteristic appearance in
time series of high-resolution spectra. Coupled to the light cycle they can be
observed blue- or red-shifted, for some phases even line doubling is found.
This is being explained by radial pulsations and shock fronts emerging in the
atmospheres. Based on dynamic model atmospheres synthetic CO line profiles were
calculated consistently, reproducing this scenario qualitatively.
\end{abstract}

\section{Molecular lines in high-resolution IR spectra of Miras}
Asymptotic Giant Branch (AGB) stars are low- to intermediate mass stars in a
late stage of their evolution. They appear as luminous late-type giants in the
HRD. Due to the large extension of their atmosphere, they have red colors and
effective temperatures of less than 3500\,K. Stars at the upper part of the AGB
are characterized by strong radial pulsations with long periods and large
amplitudes (Mira variables). The pulsation of layers inside the star leads to a
levitation of the atmosphere. In such a cool environment, molecules
and even dust can efficiently form and affect the spectra very distinctively.
The pulsation of the stellar interior triggers sound waves which become strong
radiating shock waves as they reach the outermost layers of the star. This
results in complicated structures (in density and velocity) within the extended
AGB atmospheres.

High-resolution spectroscopy in the near-IR is the major tool to study 
atmospheric kinematics of Miras (Lebzelter this volume). Numerous molecular
lines dominate the spectra of AGB stars in the NIR, where they are bright and
well observable. Originating in separated regions of various depths, lines of
different molecules and bands can be used to probe radial velocities (RV) there
by their (Doppler) shift in wavelength. CO is especially useful for these kind
of studies (\cite{HinHR82}, \cite{Nowot05a}), being abundant in all kinds of
chemistries (O-, C-rich, S stars) and present throughout AGB atmospheres and
envelopes.

\begin{figure}[]
\begin{center}
\includegraphics[width=.82\textwidth]{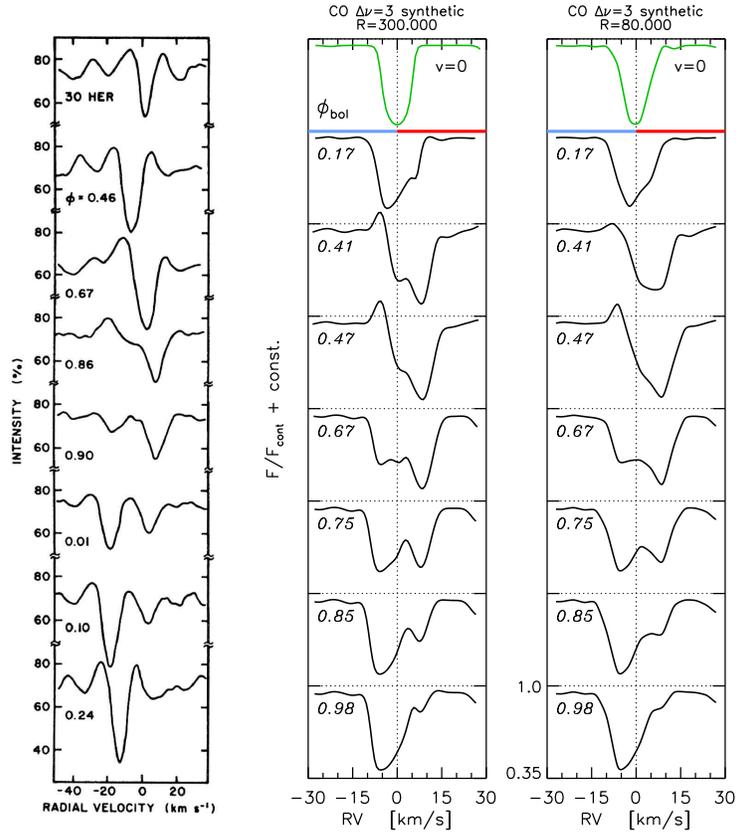}
\end{center}
\caption[]{Comparison of CO lines of the second overtone ($\Delta\,v$=3) band at
$\lambda$\,$\approx$1.6\,$\mu$m. \textit{Left:} Observed average line profiles
of $\chi$\,Cyg taken from \cite{HinHR82}, the wavelength is converted to 
velocity, superposed on observed RVs is the center of mass radial velocity
(CMRV) of the star, which is in the case of $\chi$\,Cyg --7.5\,km/s.
\textit{Right:}
Synthetic profiles for different spectral resolutions from the dynamic model
atmosphere}
\label{f:specs}
\end{figure}

By observing CO lines of the $\Delta\,v$=3 vibration--rotation bands at
$\lambda$$\approx$1.6\,$\mu$m (H-band), one can sample RVs of deep photospheric
layers in Miras (due to their excitation energies and a minimum of continous
absorption). As these layers are dominated by the pulsation of the stellar
interior, the CO second overtone lines show a characteristic behaviour.
Figure\,\ref{f:specs} (left) shows a time series of averaged line profiles for
the S-type Mira $\chi$\,Cyg of different phases during a lightcycle. 
Coupled to the variability in brightness (numbers denote visual phases with
$\phi$=0 corresponding to light maximum), periodic variations in wavelength
shift are found. While a blue-shifted line comes from outflowing matter, a
red-shifted one indicates infalling matter.\footnote{Note, that for observers
RVs are defined as being negative for outflow (towards the observer), which is
also adopted for the models for easier comparison.} Around visual maximum
even line doubling can be observed. This leads to discontinous `S-shaped'
RV-curves (e.g. Fig.\,12 of \cite{HinHR82}), usually interpreted as being due to
shock fronts running outwards through the atmosphere. For a detailed
description and interpretation of this scenario the reader is refered to
Sect.\,III. and VI.a. of \cite{HinHR82}. The results of \cite{HinSH84} and
Fig.\,1 of \cite{LebzH02} -- showing a composite of measured RVs for all Miras
studied so far -- suggest, that the shape and the amplitude
($\Delta$RV$\approx$\,25\,km/s) of this RV-curve is probably a general
characteristic of Miras, independent of their periods or spectral types.
Realistic models should therefore be able to reproduce this feature.

\begin{figure}[ht]
\begin{center}
\includegraphics[width=.7\textwidth]{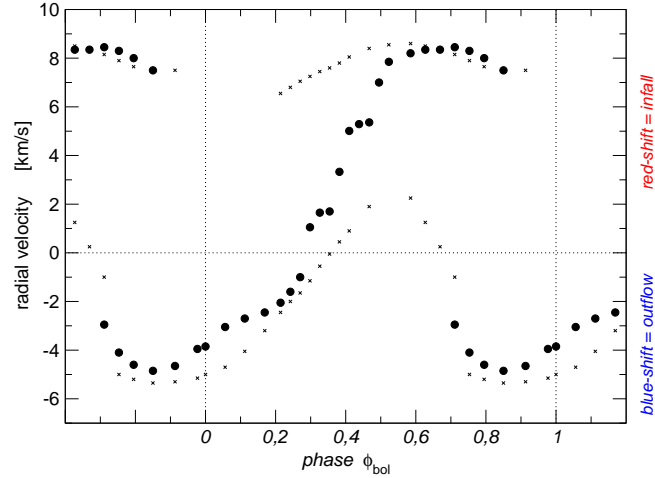}
\end{center}
\caption[]{Radial velocities derived from wavelength shifts of the synthetic CO
lines in Fig.\,\ref{f:specs}, for spectral resolutions of
$\lambda/\Delta\lambda$=300.000 (\textit{crosses}) and 70.000 (\textit{filled
circles})}
\label{f:rv}
\end{figure}

\section{Dynamic model atmospheres and synthetic line profiles}
For a realistic description of the complex and temporally variable atmospheric
structures of Miras, a combined and self-consistent solution of hydrodynamics,
frequency-dependent radiative transfer and a detailed time-dependent treatment
of dust formation is needed (H\"ofner this volume). Because of the diverse
movements of atmospheric layers in different depths, a spherical radiative
transfer including the influence of velocities on the interaction between
matter and radiation is essential to model the complex line profiles and their
variations.

High-resolution spectra were calculated to reproduce observations of the
\mbox{C-rich} Mira S\,Cep (\cite{HinkB96}), based on a dynamic model atmosphere
(as described in \cite{HoGAJ03}) with adequate stellar 
parameters.\footnote{$L_*$=10$^4$$L_{\odot}$, $M_*$=1$M_{\odot}$,
$T_*$=2600\,K, C/O=1.4, P=490\,d, $\Delta u_{\rm p}$=4\,km/s}
Details, plots of the atmospheric spatial structure and various synthetic line
profiles can be found in \cite{Nowot05a} and \cite{Nowot05b}. Since observing 
second overtone CO lines is difficult in C stars due to contamination of this 
spectral region by other molecules (CN, C$_2$), no corresponding studies have 
been published yet. Nevertheless, we tried to synthesize such line profiles 
as well, which should show a similar behaviour for all spectral types (M/S/C) 
and sample the pulsational layers like CN (see \cite{Nowot05a}, \cite{Nowot05b}). 
 The CO 5--2 P30 line from the line list of \cite{GoorC94} at 6033.8967\,cm$^{-1}$ 
was chosen for the modelling. Included in the calculations is the opacity of 
dust (pure amorphous carbon, no SiC).

Figure\,\ref{f:specs} (right) shows a time series of synthetic profiles. The
plotted phases are chosen for a direct comparison with the observed ones of the
representative Mira $\chi$\,Cyg. \textit{Bolometric} (not visual)
\textit{phases} $\phi_{\rm bol}$ are also given -- for a clear distinction
these are written \textit{italic} throughout the text. A difference of
$\approx$\,0.25 in `phase' between the two series is needed to match line
profiles of the same shape. To get an idea of the typical line width, a profile
calculated without taking velocities into account in the radiative transfer is
plotted on top. The line strengths are comparable and the synthetic
lines reproduce the typical behaviour pattern reasonably well. Line shifts are
smaller because of the lower velocities in the model, also the
doubling\footnote{Expected, as shock fronts are seen in plots of mass shells
for these dynamic models, e.g. Fig.\,2b of \cite{HoGAJ03}.} is less pronounced
(compare phases 0.01/\textit{0.75}), but still clearly visible. A blue-shifted
emission feature appearing before the line doubling phases can also be seen in
the modelled profiles (0.86/\textit{0.47}).

Figure\,\ref{f:rv} shows RVs derived from synthetic spectra of 23 calculated
phases during one lightcycle, plotted repeatedly. Line shifts were measured
from deepest points of splines fitted through the spectral points. RVs of two
other pulsational cycles duplicate these curves, they are not plotted. The
developement of a few components can be followed at higher resolution. The
splitting during phases \textit{$\approx$0.2--0.45} (which is not being
observed) turns into a continous transition from blue- to red-shifts for lower
resolution, as the two components of the line profile melt into one broad
feature (phase \textit{0.41} in Fig.\,\ref{f:specs}). Taking into account the
above mentioned shift of $\approx$\,0.25 between visual phases of observations
and \textit{bolometric} ones of the model, the resulting RV-curve compares
passably to observed ones (e.g. Fig.\,12 of \cite{HinHR82}). The `S-shape' is
reproduced as well as the asymmetry w.r.t. RV=0, line doubling appears for a
similar time interval ($\Delta\phi\approx$\,0.2). The weak component with
RV$\approx$\,0\,km/s seen around phases of \textit{0.6--0.7} at higher
resolution, can be recognised in Fig.\,1 of \cite{LebzH02} and also very
slightly in the observed $\chi$\,Cyg profile for phase 0.90.  A more detailed
analysis shows that the synthetic CO lines emerge in depth layers of
$R$=0.8--1.3\,$R_*$ with gas temperatures of $\approx$2200--3500\,K.
Although the amplitude $\Delta$RV$\approx$14\,km/s is a little too low, this
behaviour could be reproduced for the first time by consistent calculations.
For more details and results concerning other molecular lines, the reader 
is referred to \cite{Nowot05a} and \cite{Nowot05b}.



\begin{thebibliography}{8.}
\addcontentsline{toc}{section}{References}

\bibitem{GoorC94} D. Goorvitch, C.Jr. Chackerian: ApJS \textbf{91}, 483 (1994)
\bibitem{HinkB96} K.H. Hinkle, C. Barnbaum: AJ \textbf{111}, 913 (1996)
\bibitem{HinHR82} K.H. Hinkle, D.N.B. Hall, S.T. Ridgway: ApJ \textbf{252}, 697
(1982)
\bibitem{HinSH84} K.H. Hinkle, W.W.G. Scharlach, D.N.B. Hall: ApJ Suppl.
\textbf{56}, 1 (1984)
\bibitem{HoGAJ03} S. H\"ofner, R. Gautschy-Loidl, B. Aringer, U.G. J{\o}rgensen:
A\&A \textbf{399}, 589 (2003)
\bibitem{LebzH02} T. Lebzelter, K.H. Hinkle: ASP Conf. Series \textbf{259}, 556
(2002)
\bibitem{Nowot05a} W. Nowotny, B. Aringer, S. H\"ofner et al.: 
A\&A  \textbf{437}, 273 (2005)
\bibitem{Nowot05b} W. Nowotny, T. Lebzelter, J. Hron, S. H\"ofner: 
A\&A  \textbf{437}, 285 (2005)



\end{thebibliography}
\end{document}